# Static and Ultrafast Dynamics of Defects of SrTiO$_3$ in LaAlO$_3$/SrTiO$_3$ Heterostructures

X. Renshaw Wang[1,2], J. Q. Chen[1,3], A. Roy Barman[1,2], S. Dhar[1,3], Q. -H. Xu[1,4,c)], T. Venkatesan[1,2,3,b)], Ariando[1,2,a)]

[1]NUSNNI- Nanocore, National University of Singapore(NUS), Singapore 117411, [2]Department of Physics, NUS, Singapore 117542, [3]Department of Electrical and Computer Engineering, NUS, Singapre 117576, [4]Department of Chemistry, NUS, Singapore 117543

a) Electronic mail: ariando@nus.edu.sg.

b) Electronic mail: venky@nus.edu.sg.

c) Electronic mail: chmxqh@nus.edu.sg.

**A detailed defect energy level map was investigated for heterostructures of 26 unit cells of LaAlO$_3$ on SrTiO$_3$ prepared at a low oxygen partial pressure of 10$^{-6}$ mbar. The origin is attributed to the presence of dominating oxygen defects in SrTiO$_3$ substrate. Using femtosecond laser spectroscopy, the transient absorption and relaxation times for various transitions were determined. An ultrafast relaxation process of 2-3 picosecond from the conduction band to the closest defect level and a slower process of 70-92 picosecond from conduction band to intra-band defect level were observed. The results are discussed on the basis of propose defect-band diagram.**

As one of the most important oxide substrates, SrTiO$_3$ has been attracting significant research attention for the last several decades [1-7]. More recently, the observation of a quasi two dimensional electron gas (2DEG) at the LaAlO$_3$/SrTiO$_3$ interface [8-12] possibly due to polarization catastrophe [13] has triggered further attention. Consequently, a clear map of all the defect levels in LaAlO$_3$/SrTiO$_3$ heterostructures and their respective characteristics is critical for opto-electronic device applications, such as oxide light-emitting diodes, field effect transistors, solar cells and so on. It can be assumed that the optical transmission properties of a sample comprising a thin (~ 10 nm) LaAlO$_3$ layer on a SrTiO$_3$ substrate will be dominated by the substrate contribution. Recently, a picosecond (ps) pump probe study of intra-band defect levels with temporal resolution of 15 ps in SrTiO$_3$ [14] had been done. In this study, the authors pumped electrons from defect levels to the conduction band (CB) and studied the relaxation of the carriers with a probe. They measured a 40 ps relaxation time from CB to a defect level centered at 714 nm and postulated that further decay of electrons from this 714 nm defect level to another defect level below occurred in a time of more than 500 ps. However, more detailed characteristics on the dynamic properties of these defect levels could be obtained by employing a higher resolution laser pump and probe. In the present study, we report a 100 femtosecond (fs) pump probe study of LaAlO$_3$/SrTiO$_3$ heterostructures prepared by pulsed laser deposition in the oxygen pressure of $10^{-6}$ mbar. From device point of view, LaAlO$_3$/SrTiO$_3$ heterostructures prepared in the oxygen pressure range of $10^{-3}$-$10^{-4}$ mbar are interesting as they exhibit a quasi 2DEG. At lower growth pressure, the electron distribution becomes more three dimensional [15]. However, the optical properties of LaAlO$_3$/SrTiO$_3$ heterostructures prepared at oxygen pressure greater than $10^{-5}$ mbar show very few intra-band features. Therefore, our investigation is focused on heterostructures prepared in an oxygen pressure of $10^{-6}$ mbar (where oxygen defects are enhanced) from which one may obtain information for the higher growth pressures by extrapolating it.

LaAlO$_3$/SrTiO$_3$ heterostructures were prepared by depositing LaAlO$_3$ on SrTiO$_3$ (001) substrates with a TiO$_2$ terminated surface. To get such surface, SrTiO$_3$ was treated by buffered hydrogen fluoride (BHF) and annealed at 950 °C in flowing oxygen gas [5, 16]. Using pulsed laser deposition technique, 26 unit cells (uc) of LaAlO$_3$ were deposited on SrTiO$_3$ at 850 °C in oxygen partial pressure (P$_{O2}$) of 10$^{-6}$ mbar from a single-crystal LaAlO$_3$ target. The laser energy density on the target was 1.8 J/cm$^2$ and the repetition rate was 1 Hz. During deposition, layer by layer growth was monitored by *in-situ* reflection high energy electron diffraction.

The transmittance spectrum of the heterostructures was measured using a Shimadzu UV-IR spectrometer with a wavelength range from 240 to 2600 nm (Fig. 1a). As we can see from the transmittance spectrum, SrTiO$_3$ (blue curve) has an optical band gap of 3.22 eV (385 nm) and it is transparent in both visible and near-infrared regions at room temperature. However, for 26 uc of LaAlO$_3$ grown on SrTiO$_3$ in 10$^{-6}$ mbar (red curve), the transmittance spectrum drastically changed while the optical band gap remained the same.

Three absorption peaks and an infrared absorption continuum in the wavelength region above 1.2 μm are observed in the transmittance spectrum of LaAlO$_3$/SrTiO$_3$ heterostructures (Fig. 1a). The optical density (OD) plotted in Fig. 1b is computed from OD(λ) = - log$_{10}$ [T(λ)], where T(λ) is transmittance at wavelength λ normalized by the transmittance of the SrTiO$_3$ which takes reflectivity of the two surfaces into account. The infrared continuum above 1.2 μm due to free carrier absorption can be seen. In this region, the OD varies with wavelength as $\lambda^n$ Drude's Law with a fitting parameter of n = 2.61 [17].

After subtracting the free carrier absorption, three absorption peaks representing different type of defects are clearly visible in Fig. 2a. This subtracted OD was fitted by three Lorentzian functions and the corresponding fitting parameters, peak positions, peak widths, and possible origins are listed in the Table 1. Since the samples were prepared at a very low $P_{O2}$ ($10^{-6}$ mbar) and same absorption peaks has been reported [14, 18, 20], it can be assumed that the dominant defects are associated with oxygen vacancies. A schematic of our proposed defect energy level model is shown in Fig. 2b. As summarized in Table 1, the 726 nm (1.71 eV) absorption is due to the transition from the defect level at 1.51 eV to CB, the 429 nm (2.90 eV) absorption is due to the transition from valence band (VB) to the 2.90 eV defect level and the 513 nm (2.42 eV) absorption is due to the transition from the 0.80 eV defect level to the CB.

To understand the dynamic properties of these defect levels, we employed fs pump probe technique to determine the relaxation dynamics. The fs transient absorption and pump probe experiments were performed by using a Spectra-physics Ti:sapphire oscillator seeded amplifier laser system. The amplifier gives an output with a central wavelength of 800 nm, pulse duration of 100 fs and a repetition rate of 1 kHz. The output beam was split into two portions; one larger portion acted as a pump beam and the other portion was focused at a sapphire plate to generate a white light continuum (WLC). The white light beam was split into two portions: one as probe and the other as a reference to correct pulse-to-pulse intensity fluctuations. Both the pump and probe beams were focused onto the same position on the sample with diameters of about 350 and 200 μm, respectively. The delay between the pump and probe pulses was varied by a using computer-controlled translation stage. The transmittance change (ΔT/T) is detected by a photodiode and the detection wavelength is determined by a monochromator. Two types of different measurements were performed, namely single wavelength pump probe and transient absorption. In single wavelength pump probe measurements, the ΔT/T of the probe beam was detected at a single wavelength at varying delay times after excitation of the system by the pump beam. In the transient absorption measurement, the transient change of the spectra of the sample at a fixed delay time and their temporal evolution were monitored. The dynamical processes could be measured with accuracy down to ~100 fs.

Figure 3 shows the transient absorption spectra of a low $P_{O2}$ sample at different time delays after excitation at 800 nm. The linear power dependence of the signal indicated that only single photon processes dominated in these experiments. There are two well defined regimes; 450-700 nm (regime 1) and 700-775 nm (regime 2). In regime 1 the time dependent signals at different pump intensities are shown in Fig. 4a. A well defined bleaching signal was found to decay with a combination of two time constants *i.e.*, an ultra-fast component of ~1.9 ps and a dominantly slow component of 72 ps (Fig. 4b). However, in regime 2, the time dynamics are considerably different as seen in Fig. 4c. By fitting with two exponential decay functions, a well defined fast bleaching component with a time constant of 2.7 ps and a slow absorption component with a time constant of 94 ps can be clearly seen (Fig. 4d).

The initial state after the 800 nm pumping is an excited state with energetic hot electrons excited from both the 2.90 eV and the long wavelength side of 1.51 eV defect levels. After pump laser excitation, electrons thermalize (in sub-ps time) and then start decaying from CB minima to the various defect levels with different time constants [23]. The observed fast time component can be attributed to the decay from the CB to the nearest 2.90 eV defect level. The slow time constants (70-90 ps) correspond to the time constant involved in the filling of the broad 1.51 eV defect level from the CB. The depopulated 1.51 eV defect band is able to provide two processes namely, a bleaching process (which will give a positive change in the transmittance) corresponding to the excitation from 1.51 eV defect band and an induced absorption process (which will give a negative change in the transmittance) corresponding to the excitation from VB to the 1.51 eV defect band. In contrast, the 2.90 eV bands can only provide a fast bleaching process at all excitations.

In the probe regime 1, the accessible energy range is the excitation from the 1.51 and 2.90 eV bands to the CB. The excitations from the VB to available defect levels are not significant due to energy mismatch. As a result we only see a bleaching signal over this wavelength range. As can be seen in Fig. 4a, the dominant bleaching component consists of an excitation from the 1.51 eV band (on account of the large width of this band) with a decay time on the order of 70 ps, consistent with the CB to 1.51 eV band decay time. The fast bleaching component in Fig. 4a arises from the excitation from the 2.90 eV to CB since we propose the CB to these defect levels has a fast decay component (~2 ps).

In the second probe regime, the dynamics clearly shows a dominant fast bleaching at the beginning stage and later a slow absorptive component appears. The fast bleaching component can be related to the excitation from the 2.90 eV defect level and the slow absorptive component is related to the excitation from the VB to the depopulated 1.51 eV defect band (Fig. 4c). The slow time constant of 94 ps indicates the rate at which these defect levels are getting occupied by decay from the CB. The measured time of 94 ps is similar to that obtained from the first energy regime (Fig. 4d). We would expect some difference between the two time constants since in regime 1 and 2 we are measuring the time decay to different parts of the broad 1.51 eV defect level.

In summary, we report an optical study of the defect levels in $LaAlO_3/SrTiO_3$ heterostructures grown in low $P_{O2}$ of $10^{-6}$ mbar and determined the dynamical properties of these defect levels. The defects are dominated by those from the underlying $SrTiO_3$ with two sharp defect bands located at 0.8 and 2.90 eV and a broad defect band centered at 1.51 eV from the VB. By employing fs pump probe technique, we have observed an ultrafast decay which we have attributed to a transition from the CB to 2.90 eV defect level and the decay time of the transition from CB to the broad 1.51 eV defect level to be 72-94 ps.

**Acknowledgements**

We thank G. J. You, J. Huijben and H. Hilgenkamp for discussions and experimental helps and the National Research Foundation (NRF) Singapore under the Competitive Research Program 'Tailoring Oxide Electronics by Atomic Control', NUS cross-faculty grant and FRC for financial support.

**TABLE. 1.** Observed defect levels in LaAlO$_3$/SrTiO$_3$ heterostructures grown in P$_{O2}$ of 10$^{-6}$ mbar.

| Peak center (nm) | Position above VB (eV) | Position below CB (eV) | Full width half maximum (eV) | Possible origin | Ref. |
|---|---|---|---|---|---|
| 429 | 2.90 | 0.33 | 0.20 | Self-trapped exciton state/ Impurity level | [17][19][20] |
| 726 | 1.51 | 1.71 | 1.43 | Deep vacancy level, *i.e.* oxygen vacancies trapped 1 electrons | [20][21] |
| 513 | 0.80 | 2.42 | 0.58 | Impurity level / Energy level within CB | [20][22] |

**FIGURE CAPTIONS**

**FIG. 1.** (Color online) UV-VIS-IR transmittance spectra. (a) UV-VIS-IR transmittance spectra for SrTiO$_3$ (blue curve) and low P$_{O2}$ LaAlO$_3$/SrTiO$_3$ heterostructures (red curve). Three absorption peaks and a continuous absorption can be seen for low P$_{O2}$ LaAlO3/SrTiO3 heterostructures. (b) OD of low P$_{O2}$ LaAlO$_3$/SrTiO$_3$ heterostructures plotted in double log scale. A Drude fit is shown as a red straight line with fitted power n of ~2.61.

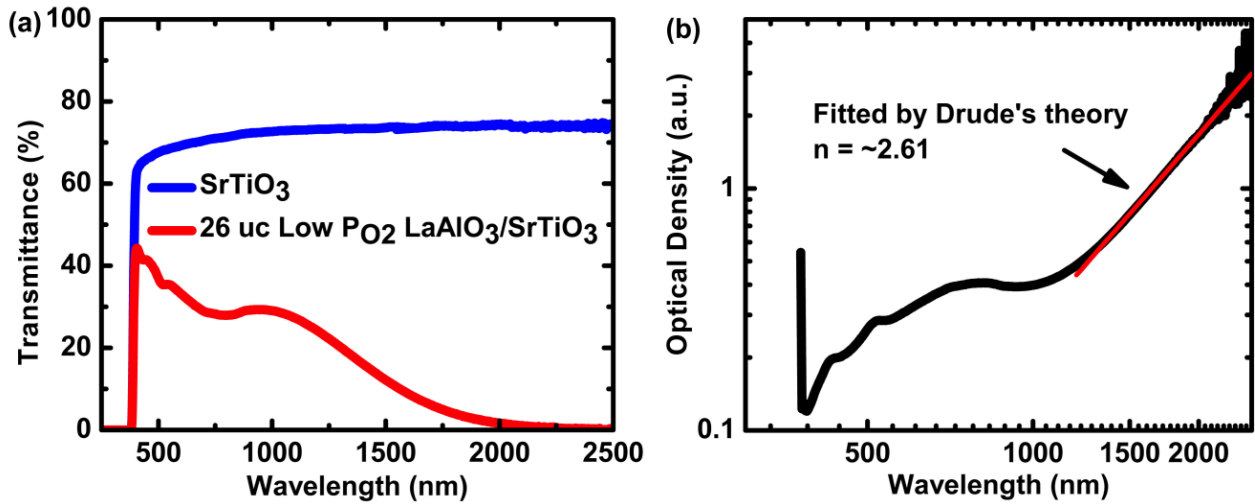

**FIG. 2.** (Color online) Defect absorption peaks and proposed energy level model. (a) Lorentzian fitting on OD of low $P_{O2}$ LaAlO$_3$/SrTiO$_3$ heterostructures. Two sharp absorption peaks and a broad absorption band are demonstrated. (b) Proposed energy level model of low $P_{O2}$ LaAlO$_3$/SrTiO$_3$ heterostructures. Defect levels drawn are not exactly to scale. For the exact full width half maximum, please refer to Table 1.

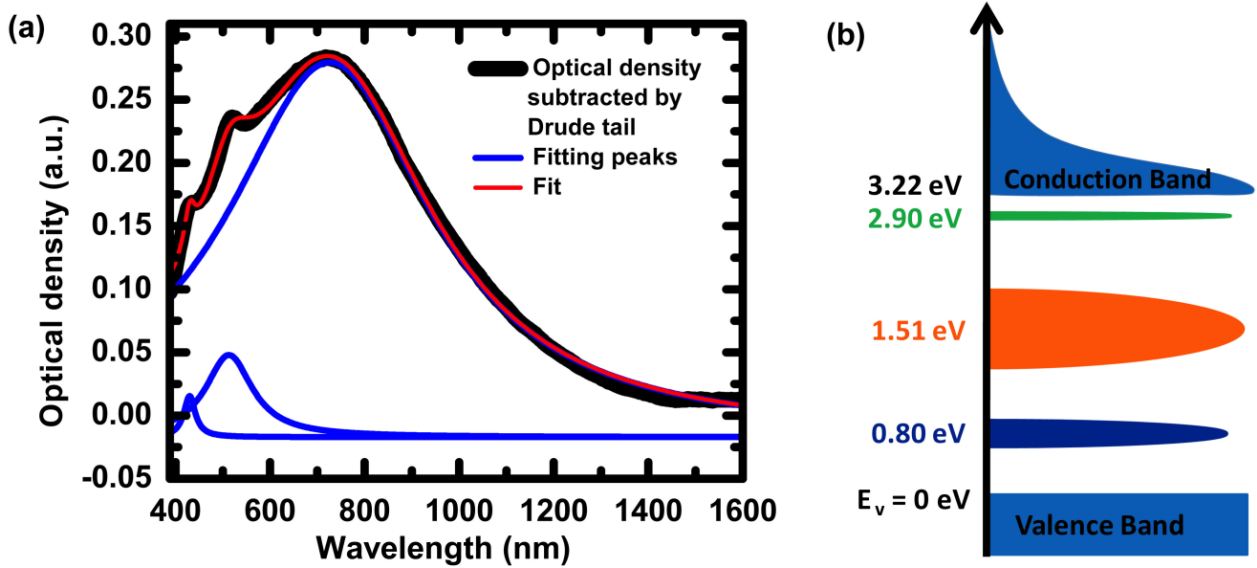

**FIG. 3.** (Color online) Transient absorption from heterostructures. Transient absorption at different delay times.

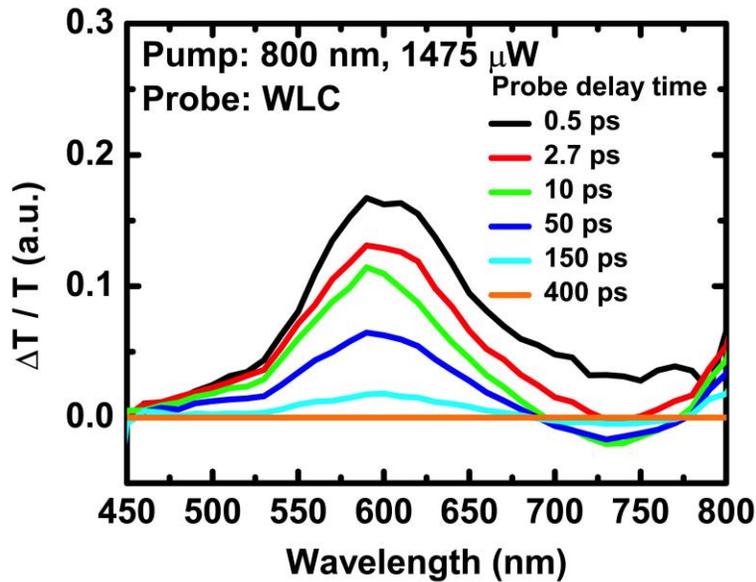

**FIG. 4.** (Color online) 800-600 nm (regime 1) and 800-750 nm (regime 2) single wavelength pump probe dynamics. (a) 800-600 nm pump probe with different pumping powers. A single photon excitation is indicated by linear function between ΔT/T and pump power. (b) 800-600 nm pump probe (on expanded time scale) with pump power of 960 μW are fitted by two fitted time constants. (c) 800-750 nm pump probe (bleaching and absorption). (d) Fitting with two exponential curves (on expanded scale) with pump power of 1311 μW.

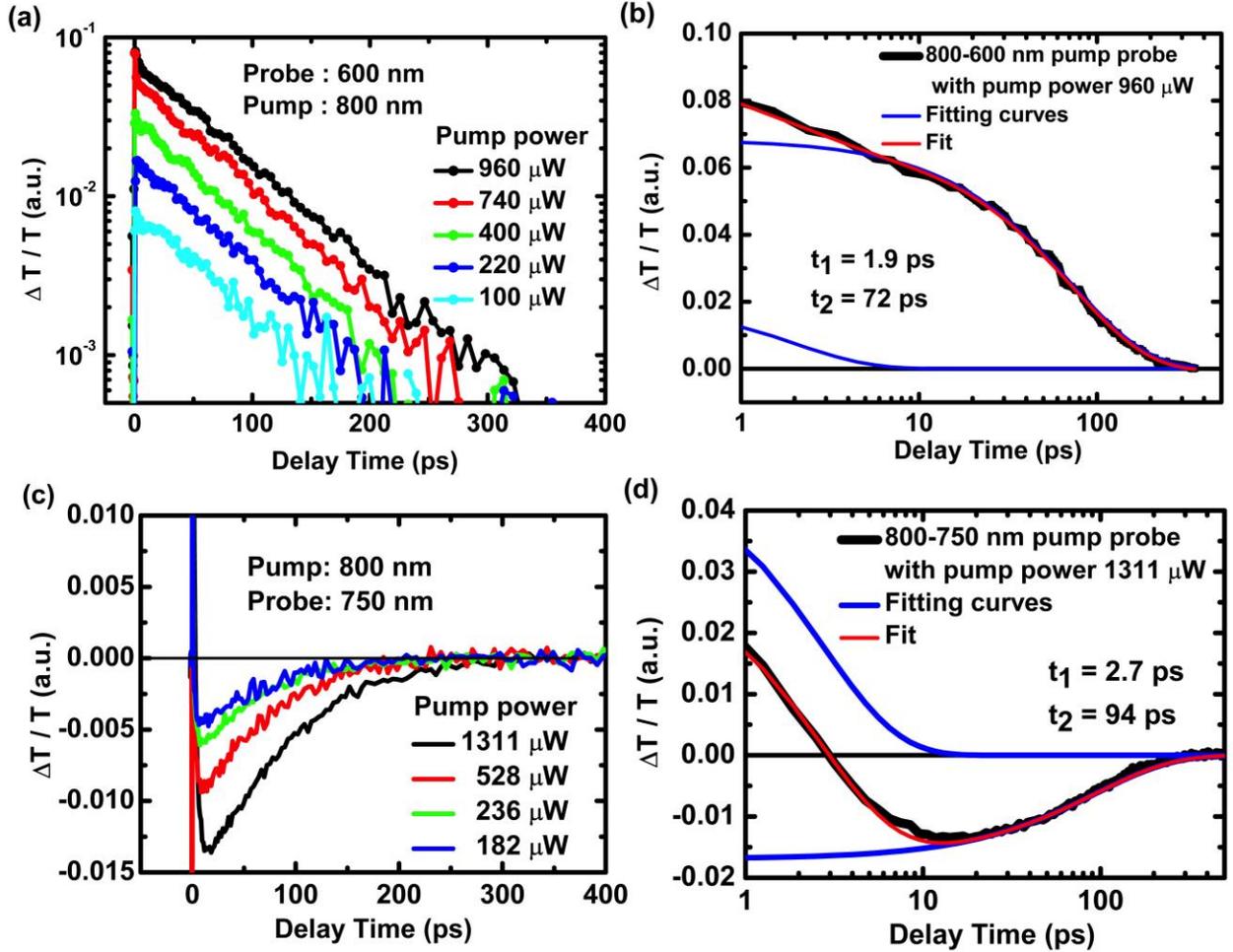